\documentclass[aps,twocolumn,nofootinbib,preprintnumbers,superscriptaddress]{revtex4-1}
\pdfoutput=1

\usepackage{tabu}
\usepackage[usenames,dvipsnames,table]{xcolor}
\usepackage{graphicx,amsmath,amssymb,amsthm,multirow,array,bm,bbm,esint}
\usepackage[mathscr]{eucal}
\usepackage[bbgreekl]{mathbbol}
\usepackage{epsf,amsfonts}
\usepackage{slashed}
\usepackage{hyperref}

\usepackage{ulem}

\hypersetup{
    pdfstartview={FitH},    
    pdftitle={title},    
    pdfauthor={authors},     
    colorlinks=true,       
    linkcolor=blue,          
    citecolor=red,        
    filecolor=magenta,      
    urlcolor=blue           
}

\newcommand{\be}{\begin{equation}}
\newcommand{\ee}{\end{equation}}
\newcommand{\bea}{\begin{eqnarray}}
\newcommand{\eea}{\end{eqnarray}}
\newcommand{\beq}{\begin{equation}}
\newcommand{\eeq}{\end{equation}}
\newcommand{\beqn}{\begin{eqnarray}}
\newcommand{\eeqn}{\end{eqnarray}}

\newcommand{\bdx}{\mathbf{x}}

\newcommand{\cL}{\mathcal{L}}

\newcommand{\zero}{{\text{\tiny (0)}}}

\usepackage{upgreek}
\newcommand{\pert}{\upvarepsilon}

\usepackage{tikz}
\usetikzlibrary{shapes.geometric, arrows}
\tikzstyle{empty} = [rectangle, minimum height=.3cm, text centered, text width=.17\textwidth, draw=black, rounded corners]
\tikzstyle{empty3} = [rectangle, minimum height=.3cm, text centered, text width=.25\textwidth, draw=black, rounded corners]
\tikzstyle{empty2} = [rectangle, minimum height=.3cm, text centered, text width=.3\textwidth, draw=black, rounded corners]
\tikzstyle{arrowT} = [very thick,<->,>=stealth']
\tikzstyle{arrow2} = [very thick,dashed,->,>=stealth']
\tikzstyle{arrowB} = [ultra thick,->,>=stealth']

\begin{document}

\title
{Higher Curvature Gravity from Entanglement in Conformal Field Theories
}

\author{Felix M. Haehl}
\author{Eliot Hijano}
\affiliation{Department of Physics and Astronomy, University of British Columbia,\\
6224 Agricultural Road, Vancouver, B.C.\ V6T 1Z1, Canada.}

\author{Onkar Parrikar}
\affiliation{David Rittenhouse Laboratory, University of Pennsylvania,\\
209 S.33rd Street, Philadelphia PA, 19104, U.S.A.}

\author{Charles Rabideau}
\affiliation{David Rittenhouse Laboratory, University of Pennsylvania,\\
209 S.33rd Street, Philadelphia PA, 19104, U.S.A.}
\affiliation{Theoretische Natuurkunde, Vrije Universiteit Brussel (VUB), and \\ International Solvay Institutes, Pleinlaan 2, B-1050 Brussels, Belgium}
%

\begin{abstract}
By generalizing different recent works to the context of higher curvature gravity, we provide a unifying framework for three related results: $(i)$ If an asymptotically AdS spacetime computes the entanglement entropies of ball-shaped regions in a CFT using a generalized Ryu-Takayanagi formula up to second order in state deformations around the vacuum, then the spacetime satisfies the correct gravitational equations of motion up to second order around AdS; $(ii)$ The holographic dual of entanglement entropy in higher curvature theories of gravity is given by Wald entropy plus a particular correction term involving extrinsic curvatures; $(iii)$ CFT relative entropy is dual to gravitational canonical energy (also in higher curvature theories of gravity). Especially for the second point, our novel derivation of this previously known statement does not involve the Euclidean replica trick.

\end{abstract}


\maketitle

\section{Introduction}

It has been known for some time that the emergence of a connected spacetime with local dynamics in holographic conformal field theories is intimately related to quantum entanglement \cite{VanRaamsdonk:2009ar,Maldacena:2013xja}. It has been understood that state dependence of relative entropy can be used to derive linearized \cite{Faulkner:2013ica} and non-linear \cite{Faulkner:2017tkh} Einstein equations. In this context, the gravitational quantity dual to relative entropy has been identified as canonical energy in Einstein gravity \cite{Lashkari:2015hha}. At the non-linear level, these statements have been established under the assumption of equal $c$- and $a$-type central charges, which is consistent with a description in terms of Einstein gravity. In this paper we aim to provide a generalization to higher curvature theories of gravity and thus for completely generic CFTs. This brings about a further interesting subtlety: while in Einstein gravity the holographic computation of CFT entanglement entropy involves evaluating the area of an appropriate extremal surface \cite{Ryu:2006bv,Hubeny:2007xt}, the entanglement entropy functional to be evaluated in higher curvature theories of gravity is more complicated \cite{Dong:2013qoa,Camps:2013zua}. Determining this functional correctly, therefore belongs to the same circle of ideas. By systematically studying higher curvature theories of gravity, we aim to provide a unifying framework for these issues. To achieve this conceptual goal, we need to overcome a technical challenge which consists of generalizing the recent CFT calculation of relative entropy \cite{Faulkner:2017tkh} to higher curvature theories of gravity.

\section{Setup}

We consider CFT states $| \psi_{\lambda} (\pert) \rangle$ that are created by sourcing the stress tensor in the Euclidean path integral over the half plane:
\begin{equation}
\label{PIstate0}
\langle \varphi_{(0)}| \psi_{\lambda} (\pert) \rangle = \int^{\varphi_{(0)}}D \varphi\, e^{- \int_{-\infty}^0 d^dx_{_E}\,\left(\cL_{CFT}+\pert\,\lambda_{\mu\nu} T^{\mu\nu}\right)}
\end{equation}
where $\varphi$ collectively denotes elementary fields in the CFT with the boundary condition $\varphi(0,\bdx) = \varphi_{(0)}(\bdx)$ at Euclidean time $x^0_E =0$. For small $\pert$ this leads to a perturbation theory in the deformation of the vacuum state. The reduced density matrix of the regioin $A$ takes the form $\rho_A = \rho_A^{(0)} + \pert \, \delta \rho_A + {\cal O}(\pert^2)$, where $\rho_A^{(0)}$ is the reduced density matrix in the vacuum.  Note that we could similarly turn on sources for other primary operators ${\cal O}_\alpha$ in the path-integral. For notational simplicity we won't do so here, but all our calculations work the same way for other primary operators.\footnote{ In fact for scalar operators, calculations are simpler since there is no bulk gauge redundancy \cite{Faulkner:2017tkh}.}

In holographic theories, this state deformation is dual to coherent excitations of the bulk gravitons.
The state perturbations in the CFT then translate to a gravitational perturbation theory of the form 
\begin{equation}
 g = g_{AdS}^{(0)} + \pert \, \delta g^{(1)} + \pert^2 \, \delta g^{(2)} + \ldots 
\end{equation}
The Einstein equations, $E^{ab} - \frac{1}{2} T^{ab} =0$, can be expanded in a similar way; $\delta^{(2)} T^{ab}$ will be quadratic in $\delta g^{(1)}$ and therefore allows us to study gravitational backreaction.

Our calculations will only be sensitive to two characteristics of the CFT: its $a$- and $c$-type central charges. We define these in general as the normalization of the universal part of ball entanglement entropy in vacuum, and the normalization of the universal stress tensor two-point function, respectively. In four-dimensional CFTs, these can be recognized as the coefficients in the conformal anomaly. The $a$-type central charge sets the AdS scale for $g_{AdS}^{(0)}$ above in the usual way.

\section{Wald formalism and beyond}
\label{sec:wald}

\begin{figure}[htbp]
\begin{center}
\includegraphics[scale=0.8]{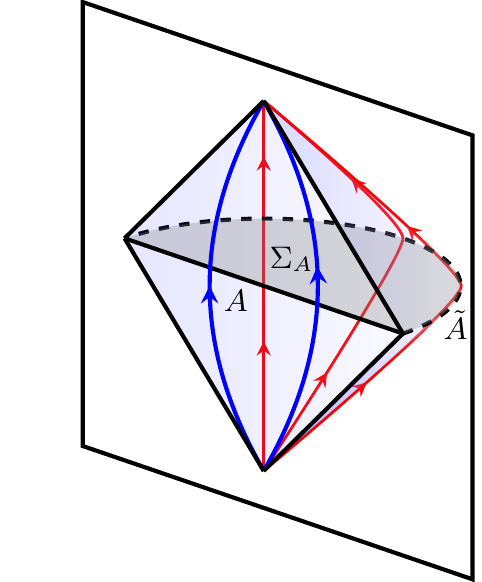}
\begin{picture}(0,0)
\setlength{\unitlength}{1cm}
\end{picture}
\caption{Associated with the ball-shaped region $A$ is a boundary domain of dependence and a Rindler wedge in the bulk with a Killing boost generator $\xi_A$ indicated in red.}
\label{fig:rindler}
\end{center}
\end{figure}

Consider a covariant theory of gravity governed by the Lagrangian $(d+1)$-form
\begin{equation}
\label{eq:fRiem}
 {\bf L} \equiv {\cal L} \, \boldsymbol{ \mathrm{vol}},\;\;\mathcal{L} = \frac{1}{16\pi G_N} \left( R + \frac{d(d-1)}{\ell^2} + f(\text{Riem}) \right) 
 \end{equation}
 where $f(\text{Riem})$ is a function of Riemann tensors, contracted with the metric. We can define the equation of motion form ${\bf E}_{_{\bf L}}$, the presymplectic potential ${\bm \theta}_{_{\bf L}}$, and the symplectic current ${\bm \omega}_{_{\bf L}}$ as
 \begin{equation}
 \label{eqn:theta_def}
 \begin{split}
 \delta {\bf L} &= - {\bf E}_{_{\bf L}} \cdot \delta g - d{\bm \theta}_{_{\bf L}}(\delta g) \,,\\
{\bm \omega}_{_{\bf L}}(\delta_1 g, \delta_2 g) & \equiv \delta_1 {\bm \theta}_{_{\bf L}}(\delta_2 g) - \delta_2{\bm \theta}_{_{\bf L}}(\delta_1 g) \,.
\end{split}
\end{equation} 
Nother's theorem can be expressed as an off-shell identity: 
\begin{equation} \label{eq:noether}
 {\bm \omega}_{_{\bf L}} (\delta g, \pounds_X g) - {\cal G}_{_{\bf L}}(\delta g,{X}) = d {\bm \chi}_{_{\bf L}} (\delta g, { X})  \,,
 \end{equation}
 where ${ X}$ is an arbitrary vector field and ${\cal G}_{_{\bf L}}$ is proportional to the equation of motion for $\delta g$. Finally, ${\bm \chi}_{_{\bf L}}$ can be defined in terms of the Noether charge, but we only require the property that its integral defines notions of modular energy and entropy. Specifically, integrating $d{\bm \chi}_{_{\bf L}}$ over a spacelike slice $\Sigma_A$ of the AdS-Rinder wedge as illustrated in Fig.\ \ref{fig:rindler}, one finds the following boundary terms:
 \begin{equation}\label{eq:gravdef}
 \begin{split}
  \int_{A} {\bm \chi}_{_{\bf L}}(\delta g, \xi_A) &= \int_A d\Sigma^\mu \,\delta T_{\mu\nu}^{grav} \, \zeta_A^\nu \equiv \delta E^{grav}_A \\
  \int_{\widetilde{A}}{\bm\chi}_{_{\bf L}}(\delta g, \xi_A) &= 8\pi \,\delta \int_{\widetilde{A}} \sqrt{\bar g} \,\frac{\partial {\cal L}}{\partial R_{abcd}}  \, n_{ab} n_{cd} \equiv \delta S^{Wald}_A
 \end{split}
 \end{equation} 
 where $\xi_A$ is the Killing vector that generates Rindler boosts, $n_{ab}$ is the binormal of the horizon and $\delta T^{grav}_{\mu\nu}$ is the holographic stress tensor, which coincides with $\delta \langle T^{CFT}_{\mu\nu}\rangle $ (see \cite{Faulkner:2013ica}). At this linear order in perturbation theory, both quantities are proportional to the ones in Einstein gravity (the proportionality constant computes the $c$-type central charge \cite{Haehl:2015rza}). Their sum can be interpreted as the Hamiltonian associated with $\xi_A$-evolution \cite{Lashkari:2016idm}.

\subsection{Second order perturbations}
 For Einstein gravity (i.e., $f(\text{Riem})=0$) Hollands and Wald \cite{Hollands:2012sf} have shown that there exists a gauge where the off-shell identity \eqref{eq:noether} holds at any order in $\pert$. Working in Gaussian null coordinates where the future- and past-directed null normal vectors to the unperturbed extremal surface are $\partial_{\ell^+}$ and $\partial_{\ell^-}$, this gauge conditions amounts to imposing: $(i)$ the Killing vector $\xi_A = 2\pi \left( \ell^+ \partial_{\ell_+} - \ell^- \partial_{\ell^-} \right)$ remains Killing near the surface, and $(ii)$ the coordinate location of the surface, defined by extremization of the entanglement entropy functional (which is Wald entropy at this order), is fixed:
 \begin{equation} \label{eq:HWgauge}
 \begin{split}
 (i)& \quad 0= \pounds_{\xi_A} g \big{|}_{\ell^+ = \ell^- = 0} \\
 (ii)& \quad 0= K^\pm_{\alpha\beta}\,\frac{\delta S_A^{\text{Wald}}}{\delta \bar{g}_{\alpha\beta}} \Big{|}_{\ell^+ = \ell^- = 0}   
 \end{split}
 \end{equation}
 For a perturbation $\delta g^{(1)}=\gamma$ in this gauge, we have the following generalization of the second order identity of \cite{Hollands:2012sf}:
 \begin{equation}
 \label{eq:HW}
 \delta^{(2)} \left[ E^{grav}_A - S^{Wald}_A \right]= \int_{\Sigma_A} {\bm \omega}_{_{\bf L}}(\gamma, \pounds_{\xi_A} \gamma) - \int_{\Sigma_A}\delta^{(2)} {\cal G}_{_{\bf L}}
 \end{equation}
where $\delta^{(2)} {\cal G}_{_{\bf L}}$ vanishes if and only if $\delta g^{(2)}$ satisfies the gravitational equations of motion associated with ${\bf L}$ at ${\cal O}(\pert^2)$.  In the higher curvature gravity context, a gauge of the form \eqref{eq:HWgauge} (such that \eqref{eq:HW} holds) exists in general. We explicitly demonstrate this for curvature squared theories by solving the corresponding inhomogeneous differential equation. The condition involving the Killing vector $\xi_A$ does not change, while the location of the surface should be defined by extremization of the promoted entanglement entropy functional \cite{Dong:2013qoa,Dong:2017xht}. Eq.\ \eqref{eq:HW} is the central gravitational identity that we use.

\subsection{Entanglement entropy functional}
\label{sec:functional}

The functional $S_A^{Wald}$ of \eqref{eq:gravdef} is a good definition of entropy for stationary Killing horizons, where various ambiguities in the definition of the Noether charge are irrelevant \cite{Wald:1993nt}. However, in a higher curvature theory of gravity and at second order in perturbation theory around AdS, it is not the correct functional for computing entanglement entropy, $S_A^{EE}$, as we will discuss now.

The Noether charge ambiguity in the definition of Wald entropy can be described according to Jacobson-Kang-Myers \cite{Iyer:1994ys,Jacobson:1993vj}. 
This JKM ambiguity can be written as a $(d-1)$-form ${\bm Y}$, which shifts the quantities entering Noether's theorem  as follows \cite{Iyer:1994ys}:
\begin{equation} \label{eq:JKMomega}
\begin{split}
  {\bm \omega}_{_{\bf L}}(\delta_1 g, \delta_2 g) &\;\;\rightarrow\;\; {\bm \omega}_{_{\bf L}} + d \big( \delta_1 {\bm Y}(\delta_2 g) - \delta_2 {\bm Y}(\delta_1 g) \big) \\
  {\bm \chi}_{_{\bf L}}(\delta g, \xi_A) &\;\;\rightarrow\;\; {\bm \chi}_{_{\bf L}}  + \delta {\bm Y}(\pounds_{\xi_A} g) - \xi_A \cdot d{\bm Y}(\delta g) 
\end{split}
\end{equation}
Note that this redefinition leaves \eqref{eq:noether} and \eqref{eq:HW} invariant.

The two extrinsic curvatures $K^\pm_{\alpha\beta}$ of $\widetilde{A}$ have boost weights $\mp 1$ (i.e., $\pounds_{\xi_A} K^\pm_{\alpha\beta} = \mp K^\pm_{\alpha\beta}$). 
Since our second order approach is only sensitive to expressions quadratic in the extrinsic curvatures, we can thus constrain the second order JKM ambiguity of the general form
\begin{equation} \label{eq:SEEansatz}
 S_A^{EE,\,\text{ansatz}} = S_A^{Wald} + \int_{\widetilde{A}} \sqrt{\bar g} \; \left( \mathfrak{B}_{\alpha\beta}^{\gamma\delta} \, K^{+\alpha\beta} \, K^-_{\gamma\delta} \right) 
\end{equation}
where $\mathfrak{B}_{\alpha\beta}^{\gamma\delta}$ is some boost-invariant tensor built out of the metric (and its derivatives). This ansatz corresponds to fixing the ambiguity as ${\bm Y}(\delta g) = {\bm Y}_{_{(EE)}} \equiv \frac{1}{2}\, {\bm \epsilon}\, \delta \big( \mathfrak{B}_{\alpha\beta}^{\gamma\delta}\, K^{+\alpha\beta} K^-_{\gamma\delta}\big)$.
Figuring out the correct entanglement entropy functional in higher derivative theories of gravity to  ${\cal O}(K^2)$, is equivalent to determining the coefficient tensor $\mathfrak{B}_{\alpha\beta}^{\gamma\delta}$ in terms of the gravitational Lagrangian. 

Once the ambiguity has been fixed, $S^{Wald}_A$ appearing in \eqref{eq:gravdef}, \eqref{eq:HWgauge} and \eqref{eq:HW} should be replaced with $S^{EE}_A$.

\subsection{Fixing the JKM ambiguity}
We wish to highlight two recent approaches to fixing this ambiguity: 
\begin{itemize}
 \item A formula for $\mathfrak{B}_{\alpha\beta}^{\gamma\delta}$ in terms of the Lagrangian can be derived using Euclidean methods and the bulk replica trick \cite{Lewkowycz:2013nqa} and demanding that entanglement entropy be computed correctly \cite{Dong:2013qoa,Camps:2013zua}. One finds (after analytic continuation):
 \begin{equation} \label{eq:extrSol}
 S_A^{EE} = S_A^{Wald} +4\pi \int_{\widetilde{A}} \sqrt{\bar g} \; \frac{\partial^2 {\cal L}}{\partial R_{+\alpha + \beta} \, \partial R_{-\gamma - \delta}} \, K^+_{\alpha\beta} \, K^-_{\gamma\delta} 
\end{equation}
 \item At second order in extrinsic curvatures, the same functional is obtained by demanding that it satisfies a linearized second law for compact horizons \cite{Wall:2015raa}.
\end{itemize}
In this paper we will derive the solution \eqref{eq:extrSol} using a third, and completely Lorentzian method. We will not use the replica trick, but instead employ a direct calculation of CFT entanglement entropy which can be matched against \eqref{eq:SEEansatz} and is a generalization of the methods recently developed in \cite{Faulkner:2017tkh} (see also \cite{Faulkner:2014jva,Faulkner:2015csl, Sarosi:2017rsq}).

Note that we only work with states which are second order in perturbation theory around AdS. Furthermore, the region $A$ for us will be ball-shaped. These are small drawbacks from the general situation (in particular, we will not encounter the ``splitting problem'' \cite{Bhattacharyya:2014yga,Miao:2014nxa} that renders the method of \cite{Dong:2013qoa} ambiguous at higher order in extrinsic curvatures), we also gain something new compared to \cite{Dong:2013qoa,Camps:2013zua}, in addition to not relying on the Euclidean replica trick: we do not need to assume large central charge or any other aspects of holography. Our results constrain a subsector of holography which is completely universal and applies to all CFTs: we will construct an entanglement entropy functional, which computes second order entanglement entropy from an auxiliary geometry for any CFT.

\subsection{Canonical energy}

We define the canonical energy in $f(\text{Riem})$ theories of gravity the same way as in Einstein gravity (as an integral over the form $\boldsymbol{\omega}_{_{\bf L}}$), but with a fixed choice of ambiguity ${\bf Y}$ in terms of extrinsic curvatures, thus generalizing \cite{Hollands:2012sf}:
\begin{equation}
\label{eq:Sgrav}
\begin{split}
&W_{\Sigma_A}(\gamma, \pounds_{\xi_A} \gamma) \equiv \delta^{(2)} \left[ E_A^{grav} - S_A^{EE} \right]  
\end{split}
\end{equation}
This is analogous to \eqref{eq:HW} after fixing the ambiguity ${\bm Y}$ in a non-trivial way as in \eqref{eq:SEEansatz}. 
We will show that the canonical energy $W_{\Sigma_A}$ is the quantity dual to relative entropy in the CFT.

\section{Relative entropy: from CFT to AdS}
\label{sec:relative}

The third ingredient for our argument is an explicit calculation of relative entropy at second order in the state perturbation \eqref{PIstate0}:
\begin{equation}\label{eq:relentdef}
\delta^{(2)} S(\rho_A||\rho_A^{(0)}) \equiv \frac{d^2}{d\pert^2} \text{Tr} \left( \rho_A \log \rho_A - \rho_A \log \rho_A^{(0)} \right) \Big{|}_{\pert=0}
\end{equation}
This quantity can only depend on the two-point function of the stress tensor used to create the state. Indeed, from \eqref{PIstate0} it is straightforward to show that the first order perturbation of the density matrix takes the form $\delta \rho_A = \rho_A^{(0)} \int \lambda_{\mu\nu} T^{\mu\nu}$, and $\delta^{(2)} S(\rho_A||\rho_A^{(0)})$ is quadratic in $\delta \rho_A$. A more careful integral expansion of the logarithm shows:
\begin{widetext}
\begin{equation}
\label{eq:calc1}
  \delta^{(2)} S(\rho_A||\rho_A^{(0)}) = - \int d\tau_a dY_a \, \widetilde\lambda_{\mu\nu}(\tau_a,Y_a) \int d\tau_b dY_b \, \widetilde\lambda_{\rho\sigma}(\tau_b,Y_b) \int_{-\infty}^\infty \frac{ds}{16\, \text{sinh}^2 \left(\frac{s+i\epsilon\, \text{sgn}(\tau_{ab})}{2} \right)}  \langle T^{\mu\nu}(\tau_a,Y_a) T^{\rho\sigma}(\tau_b+is,Y_b) \rangle 
\end{equation}
\end{widetext}
where the first two integrals are over the Euclidean space $S^1 \times \mathbb{H}^{d-1}$ with $\tau_{ab} \equiv \tau_a - \tau_b$ being the difference of Euclidean times. The sources $\widetilde\lambda_{\mu\nu}(\tau,Y) \equiv \lambda_{\mu\nu}(\tau,Y)  \Omega^{-2}(\tau,Y)$ involve a suitable conformal factor. The parameter $s$ can be thought of as (real) modular time for evolution with the modular Hamiltonian $H_A = - \log \rho_A^{(0)}$.

Importantly, the two-point function in \eqref{eq:calc1} can be represented as the asymptotic symplectic flux in an auxiliary AdS geometry. This was pointed out in \cite{Faulkner:2017tkh} for Einstein gravity. In the case of a more general gravitational Lagrangian ${\bf L}$ a systematic procedure can be followed, which is outlined in Appendix \ref{app:matching}.  
The result takes the same form as for Einstein gravity,
\begin{widetext}
\begin{equation}\label{eq:flux}
\langle T^{\rho\sigma}(\tau_a,Y_a) T^{\mu\nu}(\tau_b+is,Y_b) \rangle = - \frac{C_T}{C_T^{grav}} \int_{r_B \rightarrow \infty} ds_B \, dY_B \,  {\bm \omega}_{_{\bf L}}\big( {K_{E,{\bf L}}}_{cd}^{\rho\sigma}(is_B,r_B,Y_B|\tau_{ab},Y_a) , \, {K_{R,{\bf L}}}_{ab}^{\mu\nu}(s_B,r_B,Y_B|s,Y_b) \big) 
\end{equation}
\end{widetext}
where now ${\bf L}$ is a gravitational Lagrangian of our choice.
The gravitational computation reproduces the stress tensor correlator with a normalisation $C_T^{grav}$ fixed by the chosen gravitational Lagrangian (see Appendix \ref{app:a-c} for an explicit expression in terms of curvature squared couplings).
Since the form of two-point functions of the stress tensor is universal, 
 the correct CFT correlation function can be reproduced by any gravitational Lagrangian as long as we rescale the final answer as in \eqref{eq:flux} so as to fix the normalisation.

In order to match the CFT result for relative entropy onto the gravitational identity, the Lagrangian should be chosen such that, on the one hand, $C_T^{grav} = C_T$. Due to the universality of the two-point function of the stress tensor, $ \delta^{(2)} S(\rho_A||\rho_A^{(0)}) $ is only sensitive to the gravitational Lagrangian through this parameter. Of course, if we have a valid holographic dual we could use its bulk Lagrangian, but we wish to consider the more general possibility of constructing an auxiliary bulk theory for any CFT. 

On the other hand, the entanglement entropy for ball shaped regions in the vacuum is universal up to a normalisation $a_*$ \cite{Ryu:2006ef,Solodukhin:2008dh,Myers:2010xs,Casini:2011kv}. In order that the background geometry $g^{(0)}_{AdS}$ correctly reproduce this normalisation we must require that $a_*^{grav} = a_*$ (this fixes the parameter $\ell_{AdS}$ of our auxiliary geometry). See again Appendix \ref{app:a-c} for explicit expressions.

In \cite{Faulkner:2017tkh} it was used that for Einstein gravity, 
\begin{equation}
\label{eq:Cconstr}
a_*^{grav\,(Einstein)} = 
\frac{(d-1) \pi^d}{\Gamma(d+2)} \,
 C_T^{grav \, (Einstein)} 
 \,,
 \end{equation}
 so that the two matching conditions above imposed a CFT constraint between $C_T$ and $a_*$. For general theories of gravity,  constraint \eqref{eq:Cconstr} does not hold, so we can relax our assumptions. Instead, suitable choices of the parameters can always be found to match the $C_T$ and $a_*$ given by the CFT.

\subsection{Modular integral}
We now substitute eq. \eqref{eq:flux} in eq. \eqref{eq:calc1} for the relative entropy, and perform the integral over modular time $s$. In \cite{Faulkner:2017tkh} this integral was performed carefully for Einstein gravity. In fact, the details of this integral are the same in the present case, and so we merely state the result: 
\begin{equation} \label{eq:d2Scft}
\begin{split}
&\delta^{(2)} S(\rho || \rho_A^{(0)}) = \int_{\Sigma_A}\bm{\omega}_{_{\mathbf{L}}} \left( h, \mathcal{L}_{\xi_A}  h\right) \\
&\;\;\;
 -\int_{\mathcal{H}^+} \bm{\omega}_{_{\mathbf{L}}}\left( h,I^{(+)} \right) - \int_{\mathcal{H}^-} \bm{\omega}_{_{\mathbf{L}}} \left( h, I^{(-)} \right) \,.   
\end{split}
\end{equation}
where $\mathcal{H}^{\pm}$ are the future and past horizons of the AdS-rindler wedge, and
\beq
\begin{split}
h_{mn}(\ell^+, \ell^-, Y_B) &= \frac{1}{2} \int d\tau d Y\, \lambda_{\mu\nu}(\tau,Y) \Omega^{-2}(\tau,Y) \\
&\times K_{E;\,mn}^{\mu\nu}( \ell^+, \ell^-, Y_B | \tau, Y)
\end{split}
\eeq
is the bulk graviton sourced by the (Euclidean) boundary source. Further, $I^{(+)}$ can be conveniently expressed in terms of the contour-integral  
\begin{equation}
\begin{split}
  I^{(+)}_{mn} &= i \lim_{\ell^- \to 0} \oint_{|w| = 1-\epsilon} dw \frac{e^{-s_*^+}}{(w-e^{-s_*^+})^2} {J^p}_m{J^q}_n\\
  &\qquad\times h_{pq} \left(\frac{\ell^+}{we^{s^+_*}}, \ell^-we^{s^+_*}, Y_B\right) ,
  \end{split}
\end{equation}
where ${J^a}_b$ is the Jacobian matrix corresponding to the boost $\ell^{\pm} \to \big(we^{s_*^+}\big)^{\mp 1}\ell^{\pm}$, and similarly for $I^{(-)}_{mn}$. By expanding $h_{pq}$ around the extremal surface $\ell^{\pm} = 0$, we can easily perform the above contour integral and obtain $I^{(\pm)}_{mn}$ in an expansion around the extremal surface (see \cite{Faulkner:2017tkh} for details). It is of utmost importance to note that the metric perturbation $h_{mn}$ is in a particular gauge, namely the generalized de-Donder gauge (and not the Hollands-Wald gauge which was discussed previously). It is the nice analytic structure of the propagator in this gauge which allows us to perform the modular $s$-integral straightforwardly. 

The final step in the calculation then, is to re-express the result \eqref{eq:d2Scft} in the Hollands-Wald gauge. This is conceptually again similar to the Einstein gravity case, but in practice it is significantly more complicated. This calculation is the main new technical result of this letter, and can be found in Appendix \ref{app:calculation}. The final answer is:
\begin{equation}\label{eq:ScftHD1}
\begin{split}
\delta^{(2)} S(\rho || \rho_A^{(0)})
&= \int_{\Sigma_A} {\bm \omega}_{_{\bf L}}(g,\gamma,{\cal L}_{\xi_A} \gamma) \\
 &\;+4\pi  \int_{\widetilde{A}} \sqrt{\bar g} \; \frac{\partial^2 {\cal L}}{\partial R_{+\alpha +\beta} \, \partial R_{-\gamma-\delta}} \, \delta K^+_{\alpha\beta} \, \delta K^-_{\gamma\delta}
\end{split}
\end{equation}
where $\gamma = h + \pounds_V\, g$ is the gauge transformed perturbation such that $\gamma$ satisfies the Hollands-Wald gauge conditions \eqref{eq:HWgauge}, and the second line gives extrinsic curvature contributions which were not present in Einstein gravity. Note that \eqref{eq:ScftHD1} is an entirely field theoretic equation -- the right hand side looks like gravity, but the bulk geometry is a priori auxiliary.

\section{Results}
\label{sec:result}

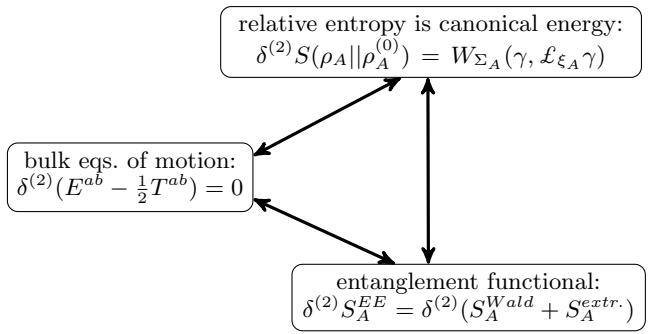
\begin{figure}
\centering
\small{
\begin{tikzpicture}[node distance=1cm]
\node (eom) [empty] {bulk eqs.\ of motion:\\ $\delta^{(2)} (E^{ab}- \frac{1}{2} T^{ab}) = 0$};
\node (canonical) [empty2, right of = eom, yshift = 1.8cm, xshift=3cm] {relative entropy is canonical energy:\\$\delta^{(2)} S(\rho_A||\rho_A^{(0)}) = W_{\Sigma_A}(\gamma, \pounds_{\xi_A} \gamma)$};
\node (functional) [empty3, right of = eom,yshift=-1.6cm, xshift=3.5cm] {entanglement functional:\\ $\delta^{(2)}S_A^{EE} = \delta^{(2)}(S_A^{Wald} + S_A^{extr.})$};
\draw [arrowT] (eom.10) -- (canonical.230);
\draw [arrowT] (eom.350) -- (functional);
\draw [arrowT] (canonical.265) -- (functional.140);
\end{tikzpicture}
}
\caption{Three concepts of interest here. We generalize all of them for higher curvature theories of gravity and show that assuming any two of them implies the third.}
\label{fig:summary}
\end{figure}

Having established the central eqs.\ \eqref{eq:HW}, \eqref{eq:Sgrav}, and \eqref{eq:ScftHD1}, we are now ready to draw some conclusions. We can solve this set of equations either for the entanglement entropy functional, or for the equations of motion, or for the canonical energy, in each case generalizing previous results, using new (Lorentzian) techniques. Taken together, these results provide a unifying framework for various ideas discussed hitherto. We now describe these three points of view. Figure \ref{fig:summary} provides a summary of these ideas.

\paragraph{Solving for the entropy functional.} We can rederive the results of \cite{Dong:2013qoa,Camps:2013zua} without any reference to Euclidean methods such as the replica trick, by solving our basic equations for the entanglement entropy functional $\delta^{(2)} S_A^{EE}$. Assuming $(i)$ the equality of canonical energy \eqref{eq:Sgrav} and boundary relative entropy \eqref{eq:ScftHD1}, and $(ii)$ the second order gravitational equations of motion (i.e., $\delta^{(2)}{\cal G}_{_{\bf L}} = 0$), we find
\begin{equation}\label{eq:Ssol}
\begin{split}
  \delta^{(2)} S_A^{EE} &=\delta^{(2)}  S_A^{Wald} \\
  &\quad -4\pi  \int_{\widetilde{A}} \sqrt{\bar g} \; \frac{\partial^2 {\cal L}}{\partial R_{+\alpha +\beta} \, \partial R_{-\gamma-\delta}} \, \delta K^+_{\alpha\beta} \, \delta K^-_{\gamma\delta}
  \end{split}
\end{equation}
This is consistent with previous results on the holographic entanglement entropy. 

\paragraph{Solving for the equations of motion.} Assuming that $(i)$ we already knew the correct entanglement entropy functional (for instance, from \cite{Dong:2013qoa,Camps:2013zua}), and $(ii)$ that again the relative entropies \eqref{eq:Sgrav} and \eqref{eq:ScftHD1} match, we could similarly solve for the integrated equations of motion: 
\begin{equation} 
\int_{\Sigma_A} \delta^{(2)} {\cal G}_{_{\bf L}} = 0 
\end{equation}
Since this holds for all ball-shaped regions (including boosted ones), one concludes ${\cal G}_{_{\bf L}} = 0$ and the equations of motion to second order in perturbation theory hold locally. This is a generalization of \cite{Faulkner:2017tkh} to the case of CFTs without any assumption about equality of $a$- and $c$-type central charges. Indeed, for any value of the central charges (such that $\frac{a-c}{c}$ is small), we can find many different bulk Lagrangians reproducing these values, and derive the associated second order equations of motion.

\paragraph{Solving for canonical energy.} Finally, we note that assuming $(i)$ the correct entanglement entropy functional, and $(ii)$ the second order equations of motion, the results \eqref{eq:HW}, \eqref{eq:Sgrav}, and \eqref{eq:ScftHD1} give a generalization of \cite{Lashkari:2015hha}: we find explicitly that 
\begin{equation}
W_{\Sigma_A}(\gamma,\pounds_{\xi_A}\gamma) = \delta^{(2)} S(\rho_A || \rho_A^{(0)})
\end{equation} 
i.e., the equality of bulk canonical energy and boundary relative entropies defined in \eqref{eq:Sgrav} and \eqref{eq:relentdef}. The new aspect with higher curvature couplings are the extrinsic curvature contributions to canonical energy, i.e., a particular fixing of the ${\bf Y}$-ambiguity in the Noether charge formalism. We have thus identified the appropriate notion of gravitational canonical energy in this scenario.

\section{Conclusion}
\label{sec:conclusions}

By studying higher curvature theories of gravity, we were able to lift some ambiguities and drop some assumptions in the context of various discussions revolving around notions of energy and entropy in gravity and CFTs. This allowed us to provide a unifying framework for understanding the intimate relations between CFT relative entropy, gravitational equations of motion, and the fixing of ambiguities in the definition of canonical energy and entanglement entropy functionals appropriate for general theories of gravity. 

Our analysis is perturbative in nature, dealing with second order deformations of the CFT vacuum state (or field excitations on empty AdS). It would be very interesting to see if the circle of ideas presented here extends to higher orders in perturbation theory. One expects that already at third order in perturbation theory, the CFT calculation will involve three-point functions. By sourcing arbitrary primary operators in the Euclidean path integral this would allow us to access the operator product expansion, and hence provide more detailed constraints based on the requirement of a geometric entropy formula. We leave this interesting question for the future.

\begin{acknowledgments}
We thank Tom Faulkner and Mark Van Raamsdonk for initial collaboration and helpful discussions. We also thank Joan Camps, Ted Jacobson, Aitor Lewkowycz, Rob Myers, Jonathan Oppenheim and Antony Speranza for helpful discussions. All of us wish to acknowledge support from the Simons Foundation through the It-from-Qubit collaboration. 
CR was also supported by the Belgian Federal Science Policy Office,
by FWO-Vlaanderen,
and by Vrije Universiteit Brussel.
\end{acknowledgments}

\appendix

\begin{widetext}

\section{Derivation of the CFT result eq.\ \eqref{eq:ScftHD1}}
\label{app:calculation}

In this appendix, we establish the central CFT result \eqref{eq:ScftHD1}.
Let us consider the various objects derived from the Lagrangian, which are relevant for our analysis. We will first reduce the general $f(\text{Riem})$ problem to the case of curvature squared theories of gravity, and then investigate the latter problem in detail. 

\subsection{Reduction to curvature squared theories}
In this subsection, we make frequent use of formulae collected in \cite{Bueno:2016ypa}, which the reader is invited to consult for more details. 

The symplectic current derived from the  $f(\text{Riem})$ part of the Lagrangian \eqref{eq:fRiem} can be written in terms of two metric perturbations $\{ \delta_1 g , \delta_2 g\}$ as follows:
\begin{equation}
\begin{split}
  {\bm \omega}^{(f)}(\delta_1 g, \delta_2 g) &= \frac{1}{16\pi G_N} \,{\bm \epsilon_c} \bigg\{ \left(S^{cabdef} \delta_1 g_{ef}+ 2 g^{gm}g^{hn} C^{cabd}_{efgh} \, \delta_1 R^{ef}{}_{mn} \right) \nabla_d \delta_2 g_{ab} \\
   &\qquad\qquad\qquad\;\; - \left( 2 \delta_1 \nabla_d P^{cabd} + g^{ef} \, \delta_1 g_{ef} \, \nabla_d P^{cabd} \right) \delta_2 g_{ab} \bigg\} -  \left[ \delta_1 \leftrightarrow \delta_2 \right] \,,
\end{split}
\end{equation}
with the following tensors defined in terms of the Lagrangian:
\begin{equation} \label{eq:Sdef}
\begin{split}
  P^{abcd} &\equiv  \frac{\partial {\cal L}}{\partial R_{abdc}}  = g^{a[c} g^{d]b} + \frac{\partial f}{\partial R_{abdc}} \,,\\
 C^{cabd}_{efgh} &\equiv g_{ek} g_{fl} g_{gm} g_{hn} \, \frac{\partial P^{cabd}}{\partial R_{klmn}}  =g_{ek} g_{fl} g_{gm} g_{hn}\frac{\partial^2 f}{\partial R_{klmn}\partial R_{cadb}} \,, \\
 S^{cabdef} &\equiv -2 P^{d(ab)(e} g^{f)c}+ 2 P^{cd(e|(a} g^{b)|f)} + P^{c(e|d(a} g^{b)| f)} + P^{c(ab)(e} g^{f)d} + P^{c(ab)d} g^{ef} \,.
\end{split}
\end{equation}
where derivatives with respect to Riemann tensor are always such that we treat the metric $g^{ef}$ and the Riemann tensor $R_{abcd}$ as independent fields and hold fixed $g^{ef}$ for purpose of the variation. For the variation of $P^{cabd}$ in the above expression, one finds 
\begin{equation}
\begin{split}
 \delta P^{cabd} 
 &= 2 g^{f[c} P^{a]ebd} \, \delta g_{ef} + g^{km} g^{ln} \, C^{cabd}_{efkl}\, \delta R^{ef}{}_{mn}  \,.
 \end{split}
\end{equation}
Now note that our computations are only sensitive to second order variations on the AdS$_{d+1}$ background. After taking the two variations, we can evaluate ${\bm \omega}^{(f)}$ on the maximally symmetric AdS geometry.
We thus find for the symplectic current:
\begin{equation} \label{eq:omegaf}
\begin{split}
  {\bm \omega}^{(f)} \Big{|}_{AdS}
   &=\frac{1}{16\pi G_N}\, {\bm \epsilon_c} \bigg\{ \left(S^{cabdef}_\zero\, \delta_1 g_{ef}+ 2 g_\zero^{gm}g_\zero^{hn} C^{\;cabd}_{\zero\, efgh} \, \delta_1 R^{ef}{}_{mn} \right) \nabla_d \delta_2 g_{ab}  - \left( 2 \delta_1 \nabla_d P^{cabd}  \right) \delta_2 g_{ab} \bigg\} -  \left[ \delta_1 \leftrightarrow \delta_2 \right]  \\
    &=\frac{1}{16\pi G_N}\, {\bm \epsilon_c} \bigg\{ \left( S^{cabdef}_\zero  + 4 g^{b[c}_\zero P_\zero^{e]afd} \right)  \delta_1 g_{ef}\nabla_d \delta_2 g_{ab} \\
    &\qquad\qquad\qquad + \left(4 g_\zero^{b[c} P_\zero^{e](ad)f} + g_\zero^{ab} P_\zero^{ecdf} - 2g_\zero^{d[c} P_\zero^{e]abf} \right) \delta_1 g_{ef}\nabla_d \delta_2 g_{ab}\\
   &\qquad\qquad\qquad + 2 g_\zero^{gm}g_\zero^{hn} C^{\;cabd}_{\zero\, efgh} \left( \delta_1 R^{ef}{}_{mn}\, \nabla_d \delta_2 g_{ab} - (\nabla_d  \delta_1 R^{ef}{}_{mn})\; \delta_2 g_{ab}  \right) \bigg\} -  \left[ \delta_1 \leftrightarrow \delta_2 \right] \,,
\end{split}
\end{equation}
where subscripts $\zero$ denote evaluation on the AdS background.

Clearly, the background tensor structures can only depend on the AdS metric.
One finds that only the following background structures are possible (i.e., compatible with the Riemann symmetries): 
\begin{equation}
\begin{split}
   P^{abcd}_\zero &=  a_0\, g^{a[c}_\zero g^{d]b}_\zero  \,,\\
   \frac{1}{\ell^2} \, C^{\;cabd}_{\zero\, efgh} &=  2 a_1 \; g^\zero_{e[g} g^\zero_{h]f} \, g_\zero^{c[b} g_\zero^{d]a}
    + 2 a_2 \; \delta^{[c}_{(m}  g_\zero^{a] [b} \delta^{d]}_{n)} \delta^m_{[e} g^\zero_{f][g}  \delta^n_{h]} + a_3 \left( \delta_e^{[c} \delta^{a]}_f \delta_g^{[b} \delta^{d]}_h +  \delta_e^{[b} \delta^{d]}_f \delta_g^{[c} \delta^{a]}_h  \right) 
\end{split}
\end{equation}
and $S^{cabdef}_\zero$ is determined in terms of $P^{abcd}_\zero$ via \eqref{eq:Sdef}.
Further, $a_{i=0,1,2,3}$ are some parameters that depend on the Lagrangian, i.e., they can be expressed in terms of $f(R_{abcd} \rightarrow - \frac{2}{\ell_{AdS}^2} g^\zero_{a[c} g^\zero_{d]b} )$ and its derivatives. The general functional $f(\text{Riem})$ is therefore reduced to merely a choice of four parameters.

For a curvature-squared Lagrangian $f_{(2)}(\text{Riem}) = \ell^2 \left( \alpha_1 \, R^2 + \alpha_2 \, R_{ab} R^{ab} + \alpha_3 \, R_{abcd} R^{abcd} \right)$, the parameters are
\begin{equation}\label{eq:match}
\begin{split}
a_0 = 1 -  2\big( d(d+1) \, \alpha_1 + d \alpha_2 + 2\,\alpha_3 \big) \frac{\ell^2}{\ell_{AdS}^2} \qquad \text{ and } \qquad
   a_i &=  \alpha_i  \quad (i=1,2,3) \,.
\end{split}
\end{equation}
It is therefore clear that curvature squared Lagrangians already contain all the parametric freedom that our second order analysis allows for: any $f(\text{Riem})$ theory is equivalent to a particular curvature squared theory for our purposes.

Using these forms of the background tensors, we can now further simplify \eqref{eq:omegaf}:
\begin{equation} \label{eq:omegaf2}
\begin{split}
  {\bm \omega}^{(f)} \Big{|}_{AdS}
   &=\frac{1}{16\pi G_N}\, {\bm \epsilon_c} \bigg\{ a_0 \, S^{cabdef}_{\text{\tiny (0,Einstein)}}   \;  \delta_1 g_{ef}\nabla_d \delta_2 g_{ab} \\
   &\qquad\quad + 2 g_\zero^{gm}g_\zero^{hn} C^{\;cabd}_{\zero\, efgh} \left( \delta_1 R^{ef}{}_{mn}\, \nabla_d \delta_2 g_{ab} - (\nabla_d  \delta_1 R^{ef}{}_{mn})\; \delta_2 g_{ab}  \right) \bigg\} -  \left[ \delta_1 \leftrightarrow \delta_2 \right] \,,
\end{split}
\end{equation}
where $S^{cabdef}_{\text{\tiny (0,Einstein)}}$ is the tensor that one obtains in Einstein gravity,
\begin{equation} \label{eq:Peinstein}
S_{\text{\tiny (0,Einstein)}}^{cabdef} =   g_\zero^{c(e}g_\zero^{f)(a} g_\zero^{b)d} - \frac{1}{2} g_\zero^{ab}g_\zero^{c(e}g_\zero^{f)d} - \frac{1}{2}  g_\zero^{c(a} g_\zero^{b)d} g_\zero^{ef}- \frac{1}{2} g_\zero^{cd} g_\zero^{e(a} g_\zero^{b)f} +\frac{1}{2} g_\zero^{ab}g_\zero^{cd}  g_\zero^{ef}
\end{equation}
If desired, one can further simplify analytically by using the identity \cite{Dong:2013qoa}
\begin{equation}
\begin{split}
 \delta R^{ef}{}_{mn} &= -  R^{ep}{}_{mn} \, g^{fq}\,\delta g_{pq}  +2\, \left[ g^{el} \, \nabla^f \nabla_m \delta g_{l n} \right]_{\text{sym}(Riem)} 
 \end{split}
\end{equation}
where sym$(Riem)$ means that the object should be symmetrized according to the symmetries of a Riemann tensor (or be contracted with an object that exhibits these symmetries, as is the case in \eqref{eq:omegaf2}). Eventually we find it more convenient to implement \eqref{eq:omegaf2} in $\mathtt{Mathematica}$. The calculations in the following subsection are otherwise tedious to perform by hand. 

From \eqref{eq:omegaf2} it is now amply clear that a calculation for curvature-squared theories of gravity completely solves our problem of general $f(\text{Riem})$ theories: the expression for ${\bm \omega}^{(f)}\big{|}_{AdS}$ decomposes into four separate terms, each of which is associated with one of the coefficients $a_{i=0,\ldots,3}$. Each of these terms leads to a calculation that is isomorphic to either Einstein gravity or one of the three curvature squared theories of gravity up to an overall factor. If one wishes to derive the second order bulk physics of some given $f(\text{Riem})$ theory, one simply needs to compute the coefficients $a_i$ for this theory and equivalently realize it through a curvature-squared theory via \eqref{eq:match}. Let us therefore now focus on curvature squared theories.

\subsection{Derivation of eq.\ \eqref{eq:ScftHD1} for curvature squared theories}
\label{app:curv_sq}

Having shown that it is sufficient to consider curvature squared theories, we will now work with the gravitational Lagrangian described by 
\begin{equation}\label{eq:Lagrangian}
{\cal L} = \frac{1}{16\pi G_N } \left( R + \frac{d(d-1)}{\ell^2} + f_{(2)}(\text{Riem}) \right) \,, \qquad f_{(2)}(\text{Riem}) = \ell^2 \left( \alpha_1 \, R^2 + \alpha_2 \, R_{ab} R^{ab} + \alpha_3 \, R_{abcd} R^{abcd} \right) 
\end{equation}
For this theory, we will now essentially repeat the calculation of \cite{Faulkner:2017tkh} with the required modifications to see the emergence of a bulk theory with higher curvature couplings. First note that the treatment of scalar fields is exactly the same as without higher derivative interactions. Let us therefore focus on the gravitational sector (i.e., states created by stress tensor deformations of the Euclidean path integral). 

We will \textit{not} assume the knowledge of the correct JKM ambiguity such that our entropy functional computes entanglement entropy. That is, we will perform the calculation from a purely CFT point of view, with no prior knowledge of how to compute entanglement entropy in the bulk, but we expect to find an entanglement functional of the form $S_A^{EE,\,\text{ansatz}}$ established in \eqref{eq:SEEansatz}. We can further simplify this problem by writing the ansatz as 
\begin{equation} \label{eq:SEEansatz2}
 S_A^{EE,\,\text{ansatz}} = S_A^{Wald} - \frac{1}{G_N}\int_{\widetilde{A}} \sqrt{\bar g} \;  \left( b_1 \, \bar{g}_{\alpha\beta}\, \bar{g}^{\gamma\delta} + b_2 \, \bar{g}_\alpha^\gamma\, \bar{g}^\delta_\beta \right) \delta K^{+\alpha\beta} \, \delta K^-_{\gamma\delta} 
\end{equation}
where the coefficients $b_{1,2}$ shall be determined in terms of the couplings $\alpha_i$.

We will now explain the strategy for deriving \eqref{eq:ScftHD1} from \eqref{eq:d2Scft}, i.e., how to go from the generalized de-Donder gauge to the Hollands Wald gauge. We start by constructing the vector field $V_{(\pm)}^a$ which generates this gauge transformation (henceforth called the ``Hollands-Wald vector field''): 
\begin{equation}
V^a = V_{(+)}^a + V_{(-)}^a \,,\qquad 
  V_{(\pm)}^a= V_{(\pm,E)}^a + \sum_{i=1}^3 \alpha_i \,  V_{(\pm,\alpha_i)}^a \,.
\end{equation} 
The labels ``$+$'' and ``$-$'' refer to the null normal coordinates $\ell_B^\pm$ transverse to the extremal surface $\widetilde{A}$. The extremal surface itself is nothing but $(d-1)$-dimensional hyperbolic space, whose coordinates we write as $x^\alpha = (u, \vec{x})^\alpha$. The full AdS metric therefore takes the form:
\beq
g^{(0)}_{AdS} = \frac{1}{4(1+\ell_+\ell_-)}\left(-\ell_-^2d\ell_+^2 - \ell_+^2d\ell_-^2\right)+\left( \frac{1}{4}+\frac{1}{4(1+\ell_+\ell_-)}\right)2 d\ell_+d\ell_- + \frac{1+\ell_+\ell_-}{u^2}\left(du^2+ d\vec{x}^2\right)\, .
\eeq
The vectors $ V_{(\pm,E)}^a $ correspond to the Hollands-Wald vector field in Einstein gravity with the following components being determined by the gauge condition \eqref{eq:HWgauge} at ${\cal O}(\alpha_i^0)$:
\begin{equation}
  \begin{split}
   V_{(\pm,E)}^\pm &= v_{(\pm,E)}^{(0)}(\ell^\mp,x^\alpha) + \ell^\pm \, v_{(\pm,E)}^{(1)} (\ell^\mp,x^\alpha) \,,\\
   V_{(\pm,E)}^\mp &= 0 \,,\\
   V_{(\pm,E)}^\alpha &= v^\alpha_{(\pm,E)}(\ell^\mp,x^\alpha) \,.
  \end{split}
\end{equation}
Further, $V_{(\pm,\alpha_i)}$ denote the corrections necessitated by higher derivative interactions such that the extremality condition associated with the entropy functional \eqref{eq:SEEansatz2} is satisfied by the perturbation $\gamma= h+{\cal L}_V g$ at $\ell^{\pm} = 0$. Explicitly, the extremality condition in the second line of \eqref{eq:HWgauge} for the functional ansatz \eqref{eq:SEEansatz2} reads 
\begin{equation}\label{eq:extremal}
\begin{split}
0&=\delta_\gamma K^\pm + 2 \ell^2 \bigg\{ -  d(d+1)\,\alpha_1 \, \delta_\gamma K^\pm   \\
&\qquad\qquad\qquad +  \left( b_1\, \nabla^2 \delta_\gamma K^\pm - \left[(d-1)b_1+d\, \alpha_2 \right] \, \delta_\gamma K^\pm \right) +\left( b_2\,\nabla_\alpha \nabla_\beta \delta_\gamma K^{\pm \alpha\beta} - \left[b_2+2\alpha_3\right] \delta_\gamma K^\pm \right) 
 \bigg\}\Big|_{\ell^{\pm}  = 0}\, .
\end{split}
\end{equation}
It turns out that this extremality condition at ${\cal O}(\alpha_i^1)$ involves only one component of the vector, which we shall dub
\begin{equation}
  V_{(\pm,\alpha_i)}^a = \delta^a_\pm \; v_{(\pm,\alpha_i)}^{(0)}(\ell^\mp,x^\alpha) \,.
\end{equation}
The calculation in this appendix does not require the ${\cal O}(\alpha_i^1)$ part of the extremality condition \ref{eq:extremal}. As we will see below, this is a direct consequence of working to linear order in $\alpha_i$. However, the ${\cal O}(\alpha_i^0)$ part of the extremality condition will be important, and simply reads
\begin{equation}\label{eq:extremal0}
0=\delta_{ h+{\cal L}_{ V_E}g} K^{\pm}\Big|_{\ell^{\pm}  = 0}\, .
\end{equation}
We do note, however, that our calculation below will fix the coefficients $b_{1,2}$ in precisely the right way such that \eqref{eq:extremal} corresponds to extremizing the correct entanglement functional appropriate for higher curvature theories.

Having established the structure of the vector field $V$, we proceed to rewrite \eqref{eq:d2Scft} in terms of $\gamma$ and $V$:
\begin{equation} \label{eq:d2Scftv2}
\begin{split}
\delta^{(2)} S(\rho || \rho_A^{(0)}) &= \int_{\Sigma_A}\bm{\omega}_{_{\mathbf{L}}} \left( \gamma, \mathcal{L}_{\xi_A}  \gamma\right) + \int_{\mathcal{H}^+}\bm{\omega}_{_{\mathbf{L}}} \left( {\cal L}_{\xi_A}\gamma, \mathcal{L}_{V}  g\right)   \\
&\quad
 -\int_{\mathcal{H}^+} \bm{\omega}_{_{\mathbf{L}}}\left( h,I(C_+)- \mathcal{L}_{[\xi_A, V_{(+)}]}g \right) - \int_{\mathcal{H}^-} \bm{\omega}_{_{\mathbf{L}}} \left( h, I(C_-) - \mathcal{L}_{[\xi_A, V_{(-)}]}g \right) \,.   
\end{split}
\end{equation}
The first term in this expression corresponds to the expected canonical energy term appearing in equation \eqref{eq:ScftHD1}. The second term can be recast as a boundary term localized at $\tilde{A}$. This is a consequence of applying equation \eqref{eq:noether} together with the linearized equations of motion obeyed by the metric fluctuation $h$, which also implies that $\mathcal{L}_{\xi_A}\gamma$ is linearly on-shell. Explicitly, the second term in \eqref{eq:d2Scftv2} is the sum of two terms, which, after integrating by parts along $\tilde{A}$ and disregarding total boundary terms, takes the form
\begin{equation}\label{eq:part1}
\begin{split}
 &\int_{\mathcal{H}^+}\bm{\omega}_{_{\mathbf{L}}} \left( {\cal L}_{\xi_A}\gamma, \mathcal{L}_{V_{(\pm)}}  g\right)  
 = \int_{{\tilde{A}}} \bm{\chi}_{_{\mathbf{L}}}\left( {\cal L}_{\xi_A}\gamma,V_{(\pm)} \right) \\
 &\quad= \frac{1}{8G_N}\int_{\tilde{A}} \sqrt{\bar{g}} \,\left\{ v^{(0)}_{(\pm,E)} {\cal X}_{(\mp,E)}  +   \sum_{i} \alpha_i  \left(   v^{(0)}_{(\pm,E)} {\cal X}_{(\mp,\alpha_i)} +2\,v^{(0)}_{(\pm,\alpha_i)}  {\cal X}_{(\mp,E)} + {\cal C}_{(\mp,\alpha_i)}  \mp 4\,h_{\bar\alpha\bar\beta}h^{\bar\alpha\bar\beta}\right)\right\}\,
\, ,
\end{split}
\end{equation}
where we have defined the following structures that are linear in $h$,
\begin{equation} 
\begin{split}
{\cal X}_{(\pm,E)} ={{\mp 2}\over{\sqrt{2}}}\delta_h K^{\pm}\, , \quad {\cal X}_{(\pm,\alpha_1)} ={\cal X}_{(\pm,\alpha_2)} =0\, ,\quad   {\cal X}_{(\pm,\alpha_3)} ={{\pm 8}\over{\sqrt{2}}}\,\delta_h\left(  K^{\pm} +\nabla_{\alpha}\nabla_{\beta} K^{\pm \alpha \beta }-\nabla_{\alpha} \nabla^{\alpha} K^{\pm}  \right)\, ,
\end{split}
\end{equation}
and the quadratic structures
\begin{equation} 
\begin{split}
{\cal C}_{(\pm,\alpha_1)} &= 0 \, ,\quad {\cal C}_{(\pm,\alpha_2)} = 0 \, ,\quad   {\cal C}_{(\pm,\alpha_3)} = \frac{\pm 8}{\sqrt{2} }\,\left( h_{\pm}^{\,\,\, \alpha} \delta_h \nabla_{\alpha} K^{\pm} -h_{\pm\alpha}\delta_h \nabla_{\beta}K^{\pm\alpha\beta} \right) \,.
\end{split}
\end{equation}
We singled out the very last term in \eqref{eq:part1} since it obviously cancels when we add the contributions of $V_{(+)}$ and $V_{(-)}$.
To write these expressions, we have made use of the linearized Einstein equations of motion (and normal derivatives thereof), viz., 
\begin{equation} 
{\cal G}_{\mu\nu}(h)=\nabla^{\rho}\nabla_{\rho} h_{\mu\nu}-\nabla_{\mu}\nabla_{\rho}h^{\rho}_{\,\,\, \nu}-\nabla_{\nu}\nabla_{\rho}h^{\rho}_{\,\,\, \mu}-2g_{\mu\nu}h +\nabla_{\mu}\nabla_{\nu}h+2h_{\mu\nu}\, .
\end{equation}
In expressions which are themselves of order ${\cal O}(\alpha_i^1)$, the two-derivative Einstein equations are sufficient and we do not keep track of corrections at higher orders in $\alpha_i$. The concrete components of the Einstein equations needed in this calculation are $\nabla_{\pm}{\cal G}^{\pm\pm}(h) $ and $ \nabla_{\mu}{\cal G}^{\mu \pm}(h)$.

The last two terms in equation \eqref{eq:d2Scftv2} can also be written as boundary terms at $\tilde{A}$. This can be seen explicitly by  integrating by parts the symplectic flux of $I(C_{\pm})-{\cal L}_{[\xi_A,V_{(\pm)}]}$ such as to isolate all dependence on $\ell^{\pm}$.  This computation turns out to isolate terms proportional to $h_{\pm\pm}\left( \ell^{\pm},x^{\alpha}  \right)\vert_{\ell^{\mp}=0}$ as  follows
\begin{equation}\label{eq:part2v0}
\begin{split}
&-\int_{\mathcal{H}^{\pm}} \bm{\omega}_{_{\mathbf{L}}}\left( h,I(C_{\pm})- \mathcal{L}_{[\xi_A, V_{(\pm)}]}g \right)  =  \frac{1}{8 G_N}\int_{\tilde{A}}\sqrt{\bar{g}}  \int d\ell^{\pm} 
\bigg\{   \mp \sqrt{2} \,   \delta_{ h+{\cal L}_{ V_{(\pm,E)}}g} K^{\pm} \\
&\qquad\qquad \mp  \left[ 
\frac{48\alpha_1 }{\sqrt{2}}\, \delta_{h+ {\cal}_{V_{(\pm,E)}}g} K^\pm +
\frac{2\alpha_2+8\alpha_3}{\sqrt{2}} \left( \nabla_\alpha \nabla^\alpha - 2 \right) \delta_{h+ {\cal}_{V_{(\pm,E)}}g} K^\pm \right]    \bigg\} h_{\pm\pm}\left( \ell^{\pm},0,x^{\alpha}  \right)  + \text{Boundary terms at }\tilde{A} 
\end{split}
\end{equation}
The terms proportional to $ \delta_{ h+{\cal L}_{ V_E}g} K^{\pm}$ vanish by virtue of the Hollands Wald condition \eqref{eq:extremal0}. In writing \eqref{eq:part2v0} we have again used the equations of motion as described above.
The  resulting boundary terms localized at $\tilde{A}$ read
\begin{equation}\label{eq:part2}
\begin{split}
& -\int_{\mathcal{H}^{\pm}} \bm{\omega}_{_{\mathbf{L}}}\left( h,I(C_{\pm})- \mathcal{L}_{[\xi_A, V_{(\pm)}]}g \right) = \frac{1}{8 G_N}\sum_{i} \alpha_i \int_{\tilde{A}(x^{\alpha})} \sqrt{\bar{g}} \left\{    v^{(0)}_{(\pm,E)} {\cal Y}_{(\mp,\alpha_i)} - v^{(0)}_{(\pm,\alpha_i)}  {\cal X}_{(\mp,E)}    \right\}   \, \\
&\qquad \qquad -\frac{1}{8 G_N} \int_{\tilde{A}(x^{\alpha})} \sqrt{\bar{g}} \left\{  2\,\alpha_2 \delta_h K^{+}\delta_h K^{-} +8\,\alpha_3 \delta_h K^{+\alpha\beta}\delta_h K^-_{\alpha\beta}  +\sum_i \alpha_i \,  {\cal C}_{(\pm,\alpha_i)} + 16 \, \alpha_3 \, h^\alpha_{[\pm} \nabla_{\mp]} \nabla_{[\alpha} h^\beta_{\beta]}   \right\}   \, ,
\end{split}
\end{equation}
where we have introduced additional structures $ {\cal Y}_{(\mp,\alpha_i)}  $ whose explicit expressions read
\begin{equation} 
\begin{split}
{\cal Y}_{(\pm,\alpha_1)} =0\, \quad {\cal Y}_{(\pm,\alpha_2)} ={{\pm 2}\over{\sqrt{2}}} \delta_h\left(  2K^{\pm}-\nabla_{\alpha}\nabla^{\alpha} K^{\pm}   \right)  \, ,\quad   {\cal Y}_{(\pm,\alpha_3)} ={{\pm 8}\over{\sqrt{2}}} \delta_h\left(  K^{\pm}-\nabla_{\alpha}\nabla_{\beta} K^{\pm \alpha\beta}   \right)\, . 
\end{split}
\end{equation}
The very last term in \eqref{eq:part2} trivially cancels once we add up the contributions from ${\cal H}^+$ and ${\cal H}^-$.
Note that equation \eqref{eq:part2} contains the familiar structures ${\cal C}_{(\pm,\alpha_i)}$ that have already appeared in formula \eqref{eq:part1}. We now replace equations \eqref{eq:part1} and \eqref{eq:part2} in \eqref{eq:d2Scftv2}  and obtain
\begin{equation} \label{eq:d2Scftv3}
\begin{split}
\delta^{(2)} S(\rho || \rho_A^{(0)}) &= \int_{\Sigma_A}\bm{\omega}_{_{\mathbf{L}}} \left( \gamma, \mathcal{L}_{\xi_A}  \gamma\right)+\frac{1}{8 G_N}\sum_{i,\pm} \alpha_i \int_{\tilde{A}(x^{\alpha})} \sqrt{\bar{g}}\,     v^{(0)}_{(\pm,E)} \left( {\cal X}_{(\mp,\alpha_i)}+{\cal Y}_{(\mp,\alpha_i)} \right) \\
&\quad- \frac{1}{8 G_N}\int_{\tilde{A}(x^{\alpha})} \sqrt{\bar{g}} \left\{  4\, \alpha_2 \,  \delta_h K^{+}\delta_h K^{-} +16\,\alpha_3\, \delta_h K^{+\alpha\beta}\delta_h K^-_{\alpha\beta}    \right\}\, .
\end{split}
\end{equation}
Note that all dependence on the ${\cal O}(\alpha_i^1)$ part of the vector $V$ has canceled out. We can further simplify this expression by using the explicit expressions of ${\cal X}_{(\pm,\alpha_i)}$ and ${\cal Y}_{(\pm,\alpha_i)}$, together with the extremality condition of equation \eqref{eq:extremal0}. The final result is
\begin{equation} \label{eq:d2Scftv4}
\begin{split}
\delta^{(2)} S(\rho || \rho_A^{(0)}) &= \int_{\Sigma_A}\bm{\omega}_{_{\mathbf{L}}} \left( \gamma, \mathcal{L}_{\xi_A}  \gamma\right)- \frac{1}{G_N}\int_{\tilde{A}(x^{\alpha})} \sqrt{\bar{g}} \left\{  \frac{\alpha_2}{2}\, \delta_{\gamma} K^{+}\delta_{\gamma} K^{-} +2\,\alpha_3\, \delta_{\gamma} K^{+\alpha\beta}\delta_{\gamma} K^-_{\alpha\beta}    \right\}\, .
\end{split}
\end{equation}
Note that the extrinsic curvature contributions are precisely the expected ones required to upgrade the Wald entropy functional to the higher curvature entanglement entropy functional. In terms of the parameters $b_i$ in the ansatz \eqref{eq:SEEansatz2} we have
\begin{equation}
b_1={{\alpha_2}\over 2}\, ,  \quad b_2=2\alpha_3\, .
\end{equation}
We conclude 
\begin{equation}\label{eq:ScftHDf}
\begin{split}
\delta^{(2)} S(\rho || \rho_A^{(0)})
= \int_{\Sigma_A} {\bm \omega}_{_{\bf L}}(g,\gamma,{\cal L}_{\xi_A} \gamma) - \delta^{(2)} S_A^{extr.} 
\end{split}
\end{equation}
One can easily check that $\delta^{(2)} S^{extr.}_A$ coincides with the extrinsic curvature terms in \eqref{eq:ScftHD1} for the curvature squared Lagrangians. This completes our derivation of the CFT identity \eqref{eq:ScftHD1}.

\subsection{$a$- and $c$-type central charges in curvature squared theories}
\label{app:a-c}
Using the parametrisation of the general curvature squared gravitational Lagrangian from Appendix \ref{app:curv_sq}, one finds \cite{Hung:2011xb} 
\begin{align}
\label{eqn:CT}
C_T^{grav} &= \frac{\Gamma(d+2) }{(d-1) \pi^\frac{d}{2} \Gamma(\frac{d}{2}) } \frac{\ell_{AdS}^{d-1}}{8\pi G_N} 
\Big[1 - 2 \big( d(d+1) \, \alpha_1 + d \alpha_2 -2(d-3)\,\alpha_3 \big) \frac{\ell^2}{\ell_{AdS}^2} \Big] \, , \\
a_*^{grav} 
&= \frac{\pi^\frac{d}{2} }{\Gamma(\frac{d}{2})} \frac{\ell_{AdS}^{d-1}}{8 \pi G_N} 
\Big[ 1 -  2\big( d(d+1) \, \alpha_1 + d \alpha_2 + 2\,\alpha_3 \big) \frac{\ell^2}{\ell_{AdS}^2} \Big] \,. 
\end{align}
If $a_*^{grav} = a_*$ is used to fix the only free parameter in the auxiliary background spacetime $g^{(0)}$ (viz., the scale $\ell_{AdS}$), then the requirement that $C_T^{grav} = C_T$ fixes the combination of the higher curvature couplings appearing in \eqref{eqn:CT} in terms of $C_T$ and $a_*$.

\section{Two-point functions as symplectic flux}
\label{app:matching}
In \cite{Faulkner:2013ica}, 
it was shown that the standard holographic dictionary for the one-point function of the stress tensor
in an arbitrary state follows from the formula for holographic entanglement entropy in general $f(\text{Riem})$ theories.\footnote{Higher curvature theories of gravity generically introduce new degrees of freedom. \cite{Faulkner:2013ica} contains some discussion of this matching. We will focus on the modes of the bulk metric which are dual to the stress tensor. The additional degrees of freedom will be treated as matter whose linearised equations of motion we will simply assume are satisfied.}
The usual statement of this dictionary is that the expectation value of an operator is given by the variation of the on-shell action by the boundary value of the dual field:
\begin{align}
\label{eqn:dictionary}
\langle T^{\mu\nu}(x) \rangle_{\psi_\lambda(\pert)} = \delta_{_{ {K_{R,{\bf L}}}_{ab}^{\mu\nu}}} \;S_{0,{\bf L}}  \,,
\end{align}
where ${K_{R,{\bf L}}}_{ab}^{\mu\nu}$ is the variation of the bulk metric at a point $X$ in the bulk under a change of the boundary metric at a point $x$ on the boundary. 
 This is given by a causal bulk to boundary propagator in the background spacetime specified by the state $\psi_\lambda(\pert)$:\footnote{See \cite{Marolf:2004fy} for a discussion of the correct causal propagator to use. That work tells us to use of the Feynman propagator when computing matrix elements of the type we need. However, we can always shift the propagator used by a linear combination of Wightman propagators, since the Wightman propagators decay sufficiently quickly near the boundary such that they do not contribute to this expression. This freedom allows us to use whichever causal propagator we wish. We will use the retarded propagator for notational continuity with our previous work \cite{Faulkner:2017tkh}.}
\begin{align}
{K_{R,{\bf L}}}_{ab}^{\mu\nu} (X|x)=\frac{ \partial g_{ab}(X)}{\partial g^\partial_{\mu\nu}(x)} \,.
\end{align}

On-shell, the variation of the action is a boundary term:\footnote{ Our notation should be understood as follows: the propagator $K_{R,{\bf L}}{}_{ab}^{\mu\nu}$ is a matrix-valued two-form. That is, its lower (bulk) indices are contracted with the measure of the integral when constructing ${\bm \theta}$, while its upper (boundary) indices survive as free inidces.}
\begin{align}
\delta_{_{ {K_{R,{\bf L}}}_{ab}^{\mu\nu}}} \;S_{0,{\bf L}}  = -\int_\partial {\bm \theta}_{_{\bf L}} \left({K_{R,{\bf L}}}_{ab}^{\mu\nu} \right) \,.
\end{align}

Taking a variation by the state on both sides and setting the reference state to the vacuum, we find  
\begin{align}
-\langle T^{\rho \sigma}(\tau_{ab},Y_a) T^{\mu \nu}(is,Y_b) \rangle 
= 
\int_{r_B\rightarrow \infty} ds_B dY_B  \; \delta_{_{{K_{E,{\bf L}}}_{cd}^{\rho\sigma}(is_B,r_B,Y_B|\tau_{ab},Y_a)}} \big[{\bm \theta}_{_{\bf L}} \left( {K_{R,{\bf L}}}_{ab}^{\mu\nu}\left(s_B,r_B,Y_B | s,Y_b \right) \right) \big]   \,.
\end{align}
We would like to write the right hand side in terms of 
${\bm \omega}_{_{\bf L}}$, but this involves a second term involving ${\bm \theta}_{_{\bf L}}$ with the order of the variations reversed. 
Since the on-shell action is a boundary term and the sources used to prepare the state must be chosen such that they do not affect the boundary conditions on the fields -- so that the prepared state is a state in the original theory we wished to consider rather than in some deformed theory -- this second term must vanish. This is the statement that 
\begin{align}
\int_{X\rightarrow\partial} dX \, {\bm \theta}_{_{\bf L}} \left(\int dx {K_{E,{\bf L}}}_{cd}^{\rho\sigma}(X|x) \lambda_{\rho\sigma}(x) \right) =0 \,,
\end{align}
in an arbitrary background state, since $ K_{E,{\bf L}}(X|x)$ itself must vanish when $X$ is taken to the boundary.

\end{widetext}


 \bibliographystyle{apsrev}
 \bibliography{inspire}

\end{document}